\documentclass[pra,twocolumn,showpacs,superscriptaddress,floatfix, reprint]{revtex4-1}

\usepackage[OT1]{fontenc}
\usepackage[usenames,dvipsnames]{color}
\usepackage[latin1]{inputenc}
\usepackage[english]{babel}
\usepackage{graphicx}
\usepackage{color}
\usepackage[Gray,squaren]{SIunits}
\usepackage{xspace}
\usepackage{ulem}
\usepackage{epstopdf}
\usepackage{amssymb,amsmath,verbatim,ulem}
\usepackage{dcolumn}
\usepackage{bm}
\usepackage{braket}
\usepackage{tabularx}
\usepackage{rotating}
\usepackage{hyperref}
\usepackage[caption=false]{subfig}

\newcommand{\RNum}[1]{\uppercase\expandafter{\romannumeral #1\relax}}

\begin{document}
\title{ Four-wave mixing based orbital angular momentum translation} 
\author{Nawaz Sarif Mallick}
\email{nawaz.phy@gmail.com}

\author{Tarak Nath Dey}
\email{tarak.dey@iitg.ac.in}
\affiliation{Department of Physics, Indian Institute of Technology
Guwahati, Guwahati, Assam 781039, India}

\date{\today}

\begin{abstract}
We theoretically study the generation of orbital angular momentum(OAM) based on four-wave mixing (FWM) process  in a diamond-type inhomogeneously broadened $^{85}$Rb atomic system.
We use density matrix formalism at weak probe limit to explain the origin of vortex translation between different optical fields and  generated signal.
We show how the singularities which are omnipresent in phase of the input optical vortex beams can be profoundly mapped to atomic coherence in the transverse plane that hold the origin of OAM translation.
This translation process works well even for moderately intense control field which enhances medium nonlinearity.
Further we have manoeuvred an additional rotation of the phase wavefront in both clockwise and anti-clockwise direction controlled by the single photon detuning.
The generation and manipulation of OAM of light beam in nonlinear medium may have important applications
in optical tweezers and quantum information processing systems.           
\end{abstract}


\maketitle
%###############################################################################%
\section{INTRODUCTION}
%###############################################################################%
Optical vortex beams carrying  singularities  in  phase  has emerged  as a topic of intensive study in the quantum domain \cite{Padgett17OAM,Arnold17OAM}. 
The azimuthal varying phase structure $e^{il\phi}$ of the beam corresponds to the origin of the orbital angular momentum \cite{ALLEN1999291}.
Various methods such as cylindrical lens pairs \cite{Padgett_2002},
computer generated hologram \cite{Heckenberg:1992}, spatial light modulator \cite{Zhu2014} ${\it etc.,}$
have been utilized to produce phase singularity.
Light beam possessing vortex singularity has broad applications in optical communication
\cite{Arnold2004,Gabriel2007,Wang2012}, super-resolution imaging \cite{Tamburini2006,Gariepy2014},
optical tweezers \cite{He1995,Neil2002,Jennifer2003}, nonlinear phenomena \cite{Nicholas2017,Leonardo2017}, ${\it etc}$.
Several systems including multi-core supermode optical fiber \cite{Li:2015,Shao2017},
photonic crystal \cite{Wang2018} and atomic vapor media \cite{Pruvost2018} are used for singularity based applications.
Specially, nonlinear optical medium has been recognized as an excellent system for studying the generation, conversion and
manipulation of vortex singularity because of its highly adaptable absorptive, dispersive and diffractive properties.

Recently, amplified spontaneous emission assisted parametric FWM processes in atomic medium has gain a lot of
attention due to its ability to translate vortex singularities from input beam to the generated beam. 
Numerous experimental studies have been performed in rubidium \cite{Arnold2012,Akulshin2015,Pruvost2018} and cesium \cite{Tabosa1999} atomic vapors for its demonstration. 
Atoms in diamond configuration have been established to efficiently map the wavefront dislocations of the near-infrared pump light to the generated collimated blue light (CBL) \cite{Arnold2012,Akulshin2015,Pruvost2018}. 
In these processes, two pump optical vortices with wavelengths
$780$ nm and $776$ nm have been involved in transferring  their OAM to forward-directed CBL with wavelength $420$ nm \cite{Arnold2012}. 
Further, this experiment has been extended to a six level system for demonstration of OAM transfer not only to CBL but also to the infra red radiation \cite{Akulshin2015}.
Recently, Chopinaud {\it et. al.,} revisited the experiment with only one single vortex pump light operating at $776$ nm \cite{Pruvost2018} and mapped high helicity vortex [-30, +30] structures into the FWM signal, which is in contrast with the previous experiments.
The theoretical counterpart of these experimental observations remains unexplored. 
A simple theory based on phase-matching condition has been adopted to explain the experimental results \cite{Arnold2012}.
A detailed study of the nonlinear atomic coherence which governs such OAM conversion processes is required. 
Further the phase wavefront dynamics of the generated beam is very essential due to the accumulation of its constituent wave vectors in different phase in presence of diffraction and dispersion of  the nonlinear atomic medium.
The competition between diffraction-induced phase and dispersion-induced phase along the transverse directions of the medium define the handedness of the singularity.
Hence a complete theoretical description of the Bloch equation for the atomic medium together with paraxial beam propagation equations for the generated beam needs to be formulated to demonstrate translation of OAM based on FWM processes.                 

In this paper, we investigate how FWM process in an inhomogeneously broadened  $^{85}$Rb atomic system can facilitate the translation of OAM associated with two optical fields to the generated field.
Two optical fields ($780$ nm and $776$ nm) and one infrared field ($5.23~\micro$m) nonlinearly
interact with the atoms and produce a non-degenerate FWM signal at 420 nm along the direction of the optical fields.
The interaction between fields and atom  form a diamond configuration.
The frequency of the generated FWM signal ($\omega_{g}$) depends on the frequency of three interacting fields and is given by $\omega_{g}=\omega_{1}+\omega_{2}-\omega_{3}$. 
The phase mismatch parameter, $\Delta \vec{k}=\vec{k}_{g}-(\vec{k}_{1}+\vec{k}_{2}-\vec{k}_{3})$ is essential to define the efficiency of the nonlinear FWM process.
The appropriate selection of propagation direction of the interacting light fields can fulfils the phase matching condition ($\Delta \vec{k}=0$).
First, we introduce density matrix formalism that enables us to derive an analytical expression for the generated atomic coherence $\rho_{41}$ in the steady state limit.
The thermal contribution of the atom can be incorporated in the coherence  by convoluting it with the thermal velocity distribution of the atom.
Next we show how the singularity which is omnipresent in phase of the probe and control fields can be efficiently mapped to $\rho_{41}$ in the transverse plane. 
This spatial inhomogeneity in absorption and phase of $\rho_{41}$ holds the key behind translation of OAM from optical fields to FWM signal.
We further  solve the Maxwell's wave equation numerically at the paraxial limit in order to delineate the successful transfer of spatial inhomogeneity from atomic coherence $\rho_{41}$ to the generated signal.
Finally, the ability to control the rotation of phase wavefront of the generated field can be achieved by changing the polarity and magnitude of the detuning of the input optical fields which is a challenging task in optical tweezers and of obvious relevance to optical trapping.

The structure of the paper is as follows. 
In section \ref{THEORETICAL}, we introduce the four-level diamond atomic system and its interaction with fields that can be described by a semi-classical density matrix formalism.
Section \ref{Perturbation} presents the analytical expression of the nonlinear coherence under weak probe approximation in order to describe the OAM translation process.
We formulate the paraxial beam propagation equation for the generated signal in section \ref{Propagation}. 
Section \ref{Vortex} provides numerical simulations which confirm the vortex translation between different optical fields to FWM signal.
Finally we briefly conclude our work in section \ref{CONCLUSION}.       
  
%###############################################################################%  
\section{THEORETICAL MODEL}
\label{THEORETICAL}
\begin{figure}
\includegraphics[scale=0.27]{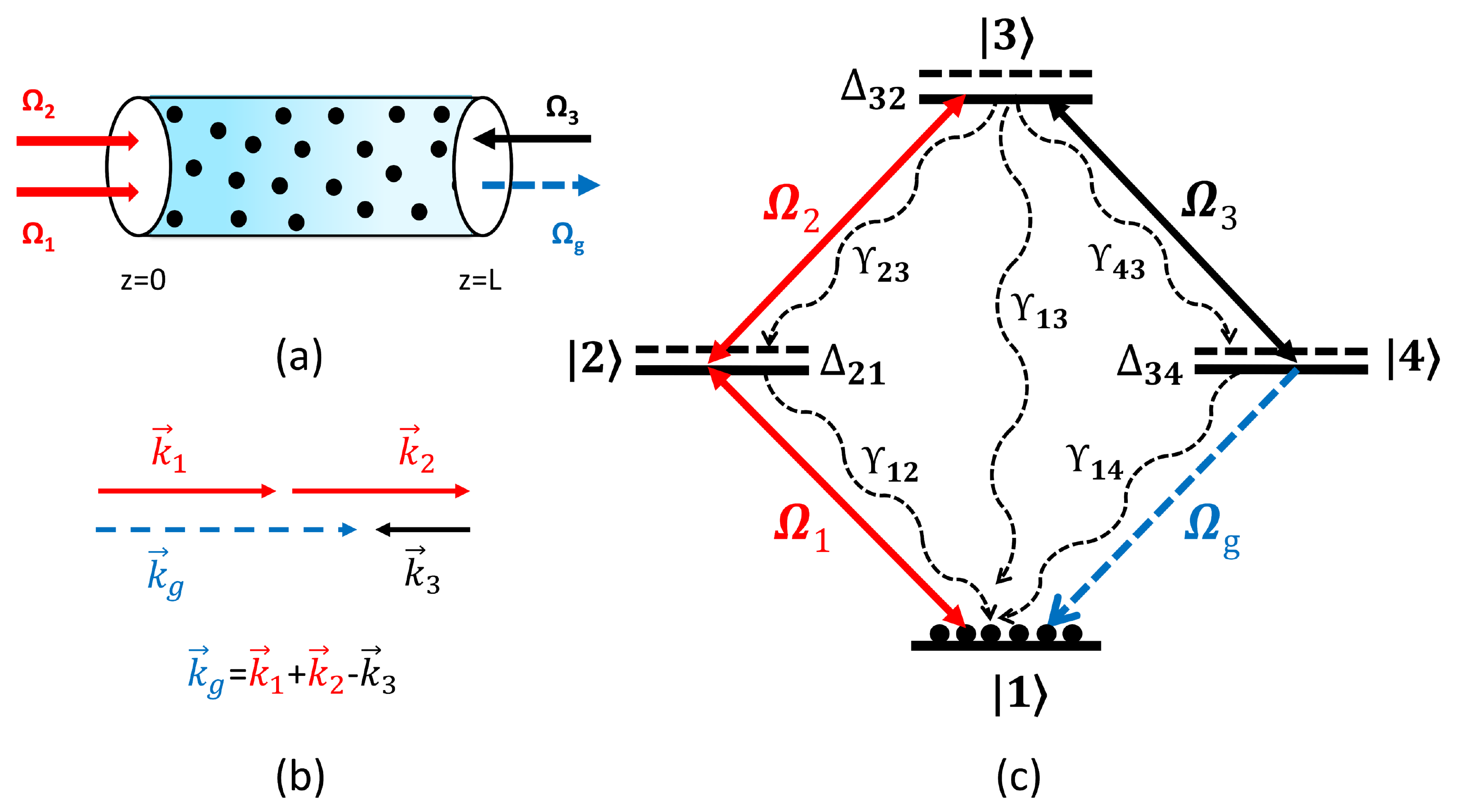}
\caption{(a) A simple block diagram of the model system. (b) The perfect phase matching configuration.
(c) Schematic representation of the four-level diamond-type atomic system. The energy states of $^{85}$Rb are defined as
$\ket1=5S_{\frac{1}{2}}$, $\ket2=5P_{\frac{3}{2}}$, $\ket3=5D_{\frac{5}{2}}$, $\ket4=6P_{\frac{3}{2}}$.}
\label{Figure1}
\end{figure}

%###############################################################################%
We consider the geometry as shown in Fig.\ref{Figure1}(a) where two co-propagating fields, namely, probe and control field, and one counter propagating infrared field interact with thermally agitated $^{85}$Rb atoms in a four-level diamond configuration.
Fig.\ref{Figure1}(b) depicts that the phase matching condition ($\Delta \vec{k}=0$) is inevitable for efficient generation of FWM signal
\cite{Model:System,Pruvost2018}.
The four-level system in Fig. \ref{Figure1}(c) consists of  three excited states $\ket 2$, $\ket 3$, $\ket 4$  and one metastable ground state $\ket 1$.
This model can be experimentally realised by considering Zeeman sublevels with  $\ket2=\ket{ 5P_{\frac{3}{2}}, F=4}$, $\ket3=\ket{5D_{\frac{5}{2}},F=5}$, $\ket4=\ket{6P_{\frac{3}{2}},F=4}$ and $\ket1=\ket{5S_{\frac{1}{2}},F=3}$ in $^{85}$Rb atomic system \cite{Model:System}.
The atomic transitions $\ket2\leftrightarrow\ket1$ and $\ket3\leftrightarrow\ket2$ are coupled by probe and control fields with wavelengths $\lambda_1=780$ nm, $\lambda_2=776$ nm whereas $\ket3\leftrightarrow\ket4$ transition is coupled by infrared field with $\lambda_3$=5.23 $\micro$m, respectively. 
All fields are defined as
%########################################
\begin{equation}\label{eq:field}
\vec{E}_{j}(\vec{r},t)=\hat e_{j}\mathcal E_{0j}(\vec{r}_{\perp})e^{i(k_j z-\omega_j t)}+c.c.,	%\quad(j\in p,c)
\end{equation}
%########################################
where $\mathcal E_{0j}(\vec{r}_{\perp})$ is the transverse variation of envelope, $k_{j}=\omega_{j}/c$ is the propagation constant, $\omega_{j}$ is the frequency and $\hat e_{j}$ is the polarisation vector of the quasi-monochromatic field.
The subscript, $j\in \{1,2,3\}$ represents the probe field, control field and infrared field.
%**************************************************************************************************************************%  
%**************************************************************************************************************************%
The interaction between the atomic transitions and electromagnetic fields can be described by the following Hamiltonian  
under electric-dipole and rotating wave approximation
%########################################
\begin{equation}\label{eq:Interaction}
\begin{aligned}
H=&\hbar \omega_{21}\ket2\bra2+\hbar(\omega_{21}+\omega_{32})\ket3\bra3\\
+&\hbar(\omega_{21}+\omega_{32}-\omega_{34})\ket4\bra4-\hbar\Omega_{1}e^{-i\omega_{1}t}\ket2\bra1\\
-&\hbar\Omega_{2}e^{-i\omega_{2}t}\ket3\bra2-\hbar\Omega_{3}e^{-i\omega_{3}t}\ket3\bra4 + h.c.,\\
\end{aligned}
\end{equation}
%########################################
where the Rabi frequencies of the probe, control and infrared field are defined as
%########################################
\begin{equation}\label{eq:Rabi}
\Omega_1=\frac{\vec{d}_{21}.\hat e_1}{\hbar} \mathcal E_{01},~~\Omega_2=\frac{\vec{d}_{32}.\hat e_2}{\hbar} \mathcal E_{02},
~~\Omega_3=\frac{\vec{d}_{34}.\hat e_3}{\hbar} \mathcal E_{03}.
\end{equation}
%########################################
The electric dipole moments $\hat {d}_{21}$, $\hat {d}_{32}$, and $\hat {d}_{34}$ allow the atomic transitions between states
$\ket2\leftrightarrow\ket1$, $\ket3\leftrightarrow\ket2$ and $\ket3\leftrightarrow\ket4$ respectively.
We now perform the following unitary transformation in order to remove explicit time dependency of the interaction Hamiltonian,
%########################################
\begin{equation}\label{eq:unitary}
H_{I}=U^\dagger H U -i\hbar U^\dagger \frac{\partial U}{\partial t},
\end{equation}
%########################################
where $U$ is defined as
%########################################
\begin{equation}\label{eq:U}
U=e^{-i\left(\omega_{1}\ket2\bra2 + (\omega_{1}+\omega_{2})\ket3\bra3 + (\omega_{1}+\omega_{2}-\omega_{3})\ket4\bra4\right)t}.
\end{equation}
%########################################
The Hamiltonian now turns into the following form
%########################################
\begin{equation}\label{eq:Hamiltonian}
\begin{aligned}
H_{I}=-&\hbar\Delta_{21}\ket2\bra2-\hbar(\Delta_{21}+\Delta_{32})\ket3\bra3\\
-&\hbar(\Delta_{21}+\Delta_{32}-\Delta_{34})\ket4\bra4-\hbar\Omega_1\ket2\bra1\\
-&\hbar\Omega_2\ket3\bra2-\hbar\Omega_3\ket3\bra4 + h.c.,\\
\end{aligned}
\end{equation}
%########################################
where the detunings are defined as
%########################################
\begin{equation}\label{eq:detuning}
\Delta_{21}=\omega_1-\omega_{21},
\Delta_{32}=\omega_2-\omega_{32},
\Delta_{34}=\omega_3-\omega_{34}.
\end{equation}
%########################################
We utilise Liouville equation to find the dynamical behaviour of the atomic populations and coherences
of the diamond-type atomic system
%########################################
\begin{equation}\label{eq:Liouville}
\dot \rho=-\frac{i}{\hbar}[H_{I},\rho]+ \mathcal{L}_\rho ,
\end{equation}
%########################################
where the second term represents various incoherent decay processes and is given by
%########################################
\begin{align}\label{decay}
\mathcal{L}_\rho &= -\sum\limits_{i=2,4} \frac{\gamma_{1i}}{2}\left(\ket i \bra i \rho-2\ket 1 \bra 1 \rho_{ii}
+\rho\ket i \bra i \right)\nonumber\\
&-\sum\limits_{j=1,2,4} \frac{\gamma_{j3}}{2}\left(\ket 3 \bra 3 \rho-2\ket j \bra j \rho_{33}+\rho\ket 3 \bra 3\right).\
\end{align}
%########################################
The spontaneous decay rates from the excited state $\ket i$, $(i \in 2,4)$ to the ground state $\ket 1$ are denoted by
$\gamma_{12}$=$\gamma_{2}$ and $\gamma_{14}$=$\gamma_{4}$.
We assume that the excited state $\ket 3$ decays to the lower state $\ket 1$, $\ket 2$ and $\ket 4$ with equal rates, {\it i.e.},
$\gamma_{13}$=$\gamma_{23}$=$\gamma_{43}$=$\gamma_{3}/3$, where $\gamma_{3}$ is the spontaneous decay rate of state $\ket 3$.
The collision rate ($\gamma_c$) and spontaneous decay rate ($\gamma_1$) of the metastable ground state, $\ket 1$ are very small and can be neglected safely. 
We substitute Eqs. (\ref{eq:Hamiltonian}) and (\ref{decay}) into the Liouville's Eq. (\ref{eq:Liouville}) and
derive the following equations of motion for the four-level atomic system :
%########################################
\begin{gather}
\begin{aligned}
\dot\rho_{11}&=\gamma_{2}\rho_{22}+\frac{\gamma_{3}}{3}\rho_{33}+\gamma_{4}\rho_{44}-i\Omega_1\rho_{12}+i\Omega^*_1\rho_{21},\\
\dot\rho_{12}&=[-i\Delta_{21}-\frac{\gamma_{2}}{2}]\rho_{12}-i\Omega_2\rho_{13}+i\Omega^*_1(\rho_{22}-\rho_{11}),\\
\dot\rho_{13}&=[-i(\Delta_{21}+\Delta_{32})-\frac{\gamma_{3}}{2}]\rho_{13}+i\Omega^*_1\rho_{23}\\
&-i\Omega^*_2\rho_{12}-i\Omega^*_3\rho_{14},\\
\dot\rho_{14}&=[-i(\Delta_{21}+\Delta_{32}-\Delta_{34})-\frac{\gamma_{4}}{2}]\rho_{14}-i\Omega_3\rho_{13}\\
&+i\Omega^*_1\rho_{24},\\
\dot\rho_{22}&=-\gamma_{2}\rho_{22}+\frac{\gamma_{3}}{3}\rho_{33}+i\Omega_1\rho_{12}-i\Omega^*_1\rho_{21}\\
&+i\Omega^*_2\rho_{32}-i\Omega_2\rho_{23},\\
\dot\rho_{23}&=[-i\Delta_{32}-\frac{\gamma_3+\gamma_2}{2}]\rho_{23}+i\Omega^*_2(\rho_{33}-\rho_{22})\\
&+i\Omega_1\rho_{13}-i\Omega^*_3\rho_{24},\\
\dot\rho_{24}&=[-i(\Delta_{32}-\Delta_{34})-\frac{\gamma_{4}+\gamma_{2}}{2}]\rho_{24}+i\Omega_1\rho_{14}\\
&+i\Omega^*_2\rho_{34}-i\Omega_3\rho_{23},\\
\dot\rho_{33}&=-\gamma_3\rho_{33}+i\Omega_2\rho_{23}+i\Omega_3\rho_{43}-i\Omega^*_2\rho_{32}-i\Omega^*_3\rho_{34},\\
\dot\rho_{34}&=[i\Delta_{34}-\frac{\gamma_{4}+\gamma_{3}}{2}]\rho_{34}+i\Omega_3(\rho_{44}-\rho_{33})+i\Omega_2\rho_{24},\\
\dot\rho_{44}&=-\gamma_4\rho_{44}+\frac{\gamma_3}{3}\rho_{33}-i\Omega_3\rho_{43}+i\Omega^*_3\rho_{34},
\end{aligned}
\label{eq:dynamical}
\raisetag{15pt}
\end{gather}
%########################################
where the overdot stands for the time derivative and star $(*)$ denotes the complex conjugate.
The diagonal density matrix elements, $\rho_{ii}$, $(i \in 1,2,3,4)$ satisfy the conservation of
population {\it i.e.}, $\rho_{11}$+$\rho_{22}$+$\rho_{33}$+$\rho_{44}$=1.
For an inhomogeneously broadened medium, thermally agitated atoms possess random finite motion which
modify the dynamics of the atomic population and coherence through Doppler frequency shift.
In presence of Doppler broadening, the probe, control and infrared field detuning can be read as $\Delta^{'}_{21}(v)=\Delta_{21}-k_{1}v$, $\Delta^{'}_{32}(v)=\Delta_{32}-k_{2}v$, and $\Delta^{'}_{34}(v)=\Delta_{34}+k_{3}v$, respectively.
The finite velocity effect can be taken into account in our simulation by averaging the coherences over the Maxwell-Boltzmann velocity distribution of the atoms as follows:    
%########################################
\begin{equation}
\langle\rho_{ij}(x,y,z)\rangle= \int \rho_{ij}(x,y,z,v)\mathcal{P}(kv)d(kv),
\end{equation}
%########################################
where the probability $\mathcal{P}(kv)d(kv)$ represent as 
%########################################
\begin{equation}\label{eq:distribution}
\mathcal{P}(kv)d(kv)=\frac{1}{\sqrt{2\pi D^{2}}}  e^{-\frac{(kv)^2}{2D^2}} d(kv),
\end{equation}
%########################################
Here the Doppler width D is given by $D=\sqrt{k_{B}T\nu^2_{c}/Mc^2}$, where $M$ is the atomic mass,
$k_{B}$ is the Boltzmann constant and $T$ is the thermal equilibrium temperature. 
For $^{85}$Rb atoms, $D$ is 37$\gamma$ at room temperature (T=300K) \cite{Sell:OL14}.

%###############################################################################%
\subsection{Perturbative analysis}
\label{Perturbation}
%###############################################################################%
In this section, we derive an analytical expression for the atomic coherence under 
$\Omega_1\ll\Omega_2, \Omega_3$ that asserts the probe field ($\Omega_1$) to be treated as a perturbation to the system.
This  analysis is valid for all orders for the control and infrared fields.
Now the time independent solutions of the density-matrix equations can be expressed as
%########################################
\begin{equation}\label{eq:perturbation1}
\begin{aligned}
\rho_{ij}&=\rho_{ij}^{(0)}+\Omega_1\rho_{ij}^{(1)}+\Omega_1^* \rho_{ij}^{(2)}\\
\end{aligned}
\end{equation}
%########################################
where $\rho_{ij}^{(0)}$ is the solution in the absence of the probe field and $\rho_{ij}^{(k)}$, $k\in \{1,2\}$
describes the solution at positive and negative frequencies of the probe field, respectively.
We now substitute Eq. (\ref{eq:perturbation1}) in Eqs. (\ref{eq:dynamical}) by considering time derivative to be zero.
We obtain two sets of ten coupled linear equations by equating the coefficients of $\Omega_1$ and $\Omega_1^*$. 
Next, we solve these algebraic equations to
derive the atomic coherences $\rho_{41}$. 
The steady-state value of the atomic coherence, $\rho_{41}$ can be expressed
by the following expression
%########################################
\begin{equation}\label{eq:perturbation}
\rho_{41}=\frac{i\Omega_1 \Omega_2 \Omega^{*}_3 }{\Gamma_{41} [\Gamma_{31}+\frac{|\Omega_3|^2}{\Gamma_{41}}] [\Gamma_{21}+\frac{|\Omega_2|^2}{\Gamma_{31}+\frac{|\Omega_3|^2}{\Gamma_{41}}}]},\\
\end{equation}
%########################################
where
%########################################
\begin{equation}
\begin{aligned}
&\Gamma_{21}=i\Delta_{21}-\frac{\gamma_2 + \gamma_1}{2},\\
&\Gamma_{31}=i(\Delta_{21}+\Delta_{32})-\frac{\gamma_3 + \gamma_1}{2},\\
&\Gamma_{41}=i(\Delta_{21}+\Delta_{32}-\Delta_{34})-\frac{\gamma_4 + \gamma_1}{2}.\\
\end{aligned}
\end{equation}
%########################################
The numerator of Eq. (\ref{eq:perturbation}) clearly shows that the structure of the generated atomic coherence can be  modulated along the azimuthal plane by selecting appropriate phase and intensity  profiles of probe and control fields.
Furthermore, the denominator of Eq. (\ref{eq:perturbation}) can provide additional phase to the phase profile of the generated beam.
The rotation of the generated wavefront can be controlled by changing the detuning values from resonance to  off-resonance.
This allows us efficient generation and manipulation of OAM of the  optical beam which may have important applications in information science and optical communication.

%###############################################################################%
\subsection{Beam propagation equation}
\label{Propagation}
%###############################################################################%
The study of Maxwell's wave equations is essential to demonstrate how the generated atomic coherence $\rho_{41}$ can  produce a light beam with desired OAM and rotation.
The wave equation for the generated light beam can be written as
%########################################
\begin{equation}\label{eq:wave_eq}
\left(\nabla^2+\frac{1}{c^2}\frac{\partial^2}{\partial t^2}\right)\vec{E}_g=\frac{4\pi}{c^2}\frac{\partial^2 \vec{\mathcal P}_g}{\partial t^2},
\end{equation}
%########################################
where $\vec{E}_g$ is the electric field of the generated signal and $\vec{\mathcal P}_g$ is the induced polarisation due to the
probe, control and infrared fields.
The induced macroscopic polarizations can be expressed in terms of the atomic coherence as
%########################################
\begin{equation}\label{eq:polarization}
\vec{\mathcal P}_g={\mathcal N}\left(\vec{d}_{14}\rho_{41}e^{-i\omega_g t}+c.c.\right)
\end{equation}
%########################################
Under slowly varying envelope and paraxial wave approximation, the wave Eq. (\ref{eq:wave_eq}) can be cast into the following form
%########################################  
\begin{equation}\label{propagation}
\begin{aligned}
\frac{\partial \Omega_g}{\partial z}=\frac{i}{2{k_g}} \nabla_{\bot}^{2}\Omega_g + i\eta \langle\rho_{41}\rangle,~~\Omega_g=\frac{\vec{d}_{41}.\vec{\mathcal E}_{g}}{\hbar} 
\end{aligned}
\end{equation}
%########################################

The first order derivative with respect to $z$ on the left hand side indicates the variation of the amplitude of the generated signal envelope, $\Omega_g$ along the length of  the medium. 
The first term on the right hand side represents the beam's phase induced diffraction and the rotation of the wave front during its propagation. 
The last term on the right-hand side accounts for the generation and dispersion
of the medium. The coupling constant, $\eta$, is defined as $\eta=3 \mathcal N \gamma \lambda^2/8 \pi$.
The atomic density, $\mathcal N$ is $5\times 10^{12}$ atoms/cm$^3$ \cite{Akulshin2015,Pruvost2018} and
wavelength of the FWM signal, $\lambda$ is 420 nm. 
We use appropriate spatially dependent transverse profile of the optical fields to generate
the FWM signal. For this purpose, the transverse spatial profile of the optical beams
are taken to be of Laguerre-Gaussian mode $(m)$ with topological charge $l$ that can be expressed as
%########################################
\begin{gather}
\begin{aligned}
\Omega_{j}(r,\phi,z)&=\Omega^{0}_{j} \frac{w_{j}}{w_{j}(z)} \left(\frac{r\sqrt{2}}{w_{j}(z)}\right)^{\left|l\right|}
e^{-\frac{r^{2}}{w^{2}_{j}(z)}} L^{l}_{m}\left[\frac{2r^{2}}{w^{2}_{j}(z)}\right]\\
&\times e^{il\phi}e^{\frac{ik_{j}r^{2}}{2R(z)}} e^{-i(2m+|l|+1)\tan^{-1}(\frac{z}{z_0})},\\
&r=\sqrt{x^2+y^2},~~~\phi=\tan^{-1}\left(\frac{y}{x}\right),\\
&\psi(m,l,z)=(2m+|l|+1)\tan^{-1}(z/z_0).
\end{aligned}
\label{control} 
\raisetag{15pt}
\end{gather}
%######################################## 
where $\Omega^{0}_{j}$ is the input amplitude, $R(z)=z+z^{2}_0/z$ is the radius of curvature and
$z_0=\pi w^{2}_{j}/\lambda_s$ is the Rayleigh length of the beam. The spot size of the beam is defined
as $w_{j}(z)=w_{j}\sqrt{1+(z/z_0)^2}$, where $w_{j}$ is the minimum beam waist at $z$=0.
The subscript, $j\in \{1,2\}$ refers to the probe and control fields.

\section{Vortex beam generation}
\label{Vortex}
\begin{figure}[h]
\begin{center}
\includegraphics[scale=0.45]{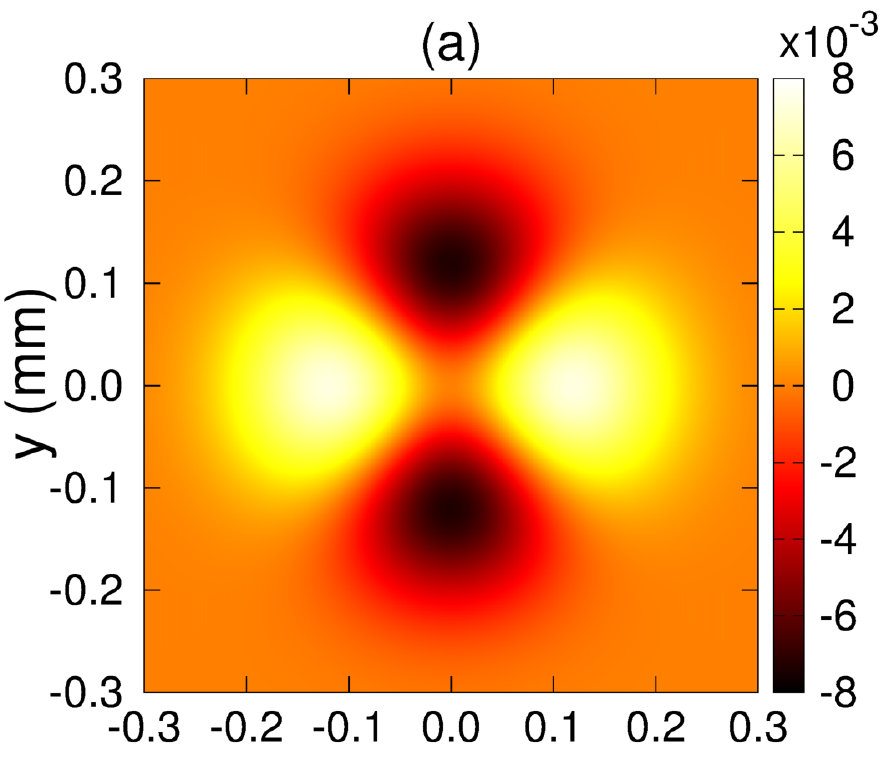}
\includegraphics[scale=0.45]{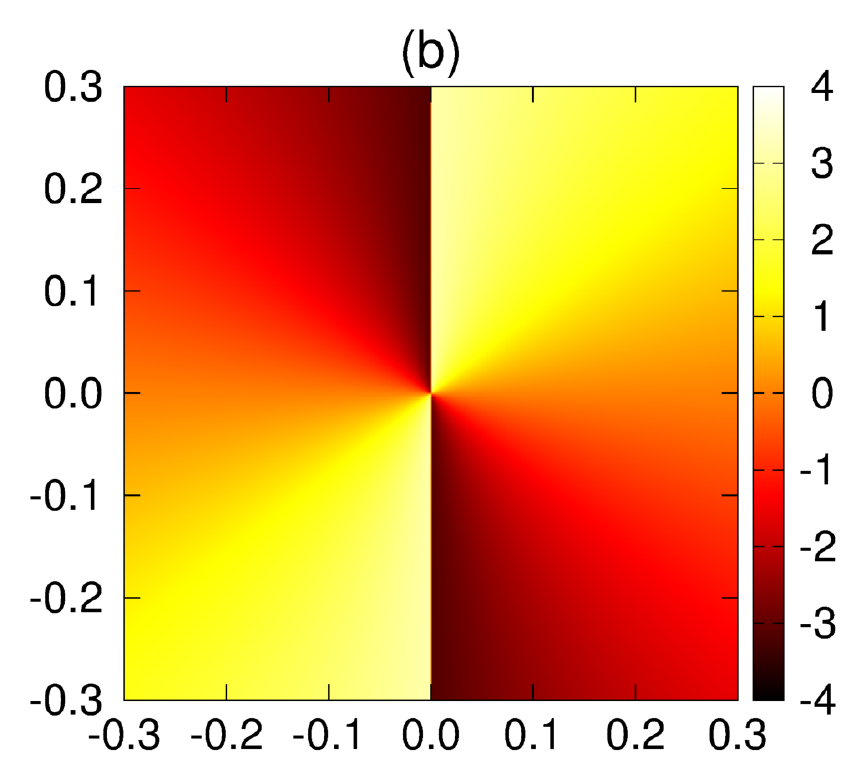}
\includegraphics[scale=0.45]{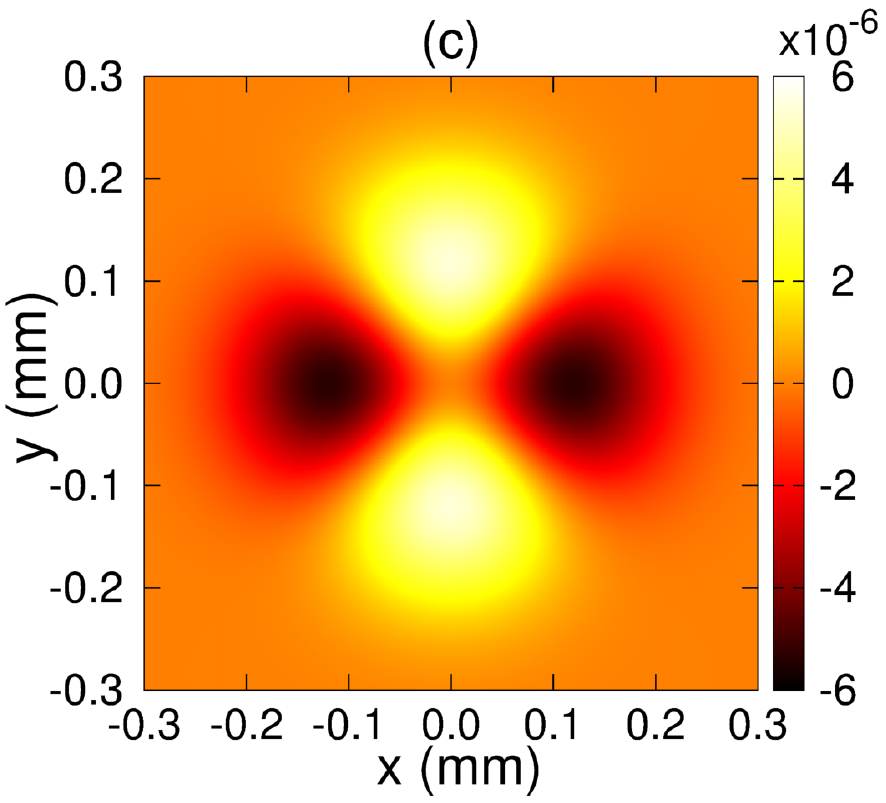}
\includegraphics[scale=0.45]{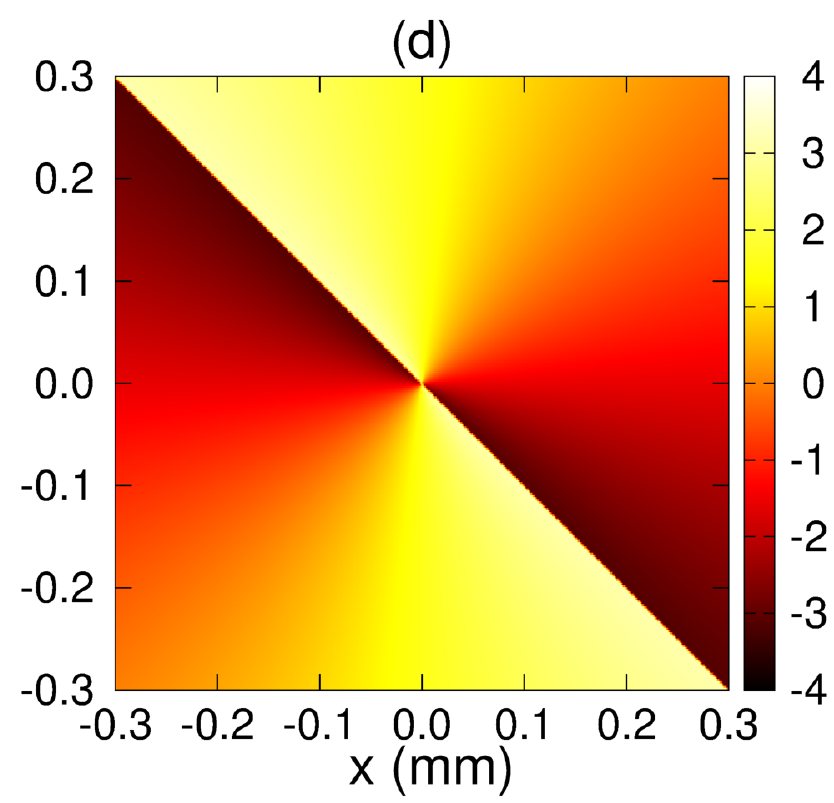}
\caption{(a) Amplitude (Re[$\Omega_1$]) and (b) phase structure of Laguerre-Gaussian probe beam.
(c) Absorption (Im[$\langle\rho_{41}\rangle$]) and (d) phase profile of the Doppler averaged atomic coherence, $\langle\rho_{41}\rangle$
in the transverse plane when $\Omega_1$ possesses the optical vortex.
The parameters are $m_{1}=0$, $l_{1}=2$, $w_{1}=120\mu m$, $\Omega^{0}_{1}=0.01\gamma$, $\Omega^{0}_{2}=0.1\gamma$,
$\Omega^{0}_{3}=1.0\gamma$, $\Delta_{21}=0$, $\Delta_{32}=0$, $\Delta_{34}=0$, T=300K.}
\label{Figure2i}
\end{center}
\end{figure}

%###############################################################################%
The aim of this section is to provide a detail theoretical explanation behind  vortex beam generation. 
First we show how input singular beams generate spatial inhomogeneity in absorption and singularity in phase of $\rho_{41}$ that hold the key behind translation of OAM from input optical fields to FWM signal.
In this regard, we integrate $\rho_{41}$ (\ref{eq:perturbation}) by taking into account the Doppler velocity distribution (\ref{eq:distribution}).
In order to satisfy the perturbation condition, we chose the control field ($\Omega_{2}=0.1\gamma$) and infrared field ($\Omega_{3}=\gamma$) intensities to be much larger than the probe field ($\Omega_{1}=0.01\gamma$) intensity.
We have chosen the probe field to be a singular beam while keeping the rest of the two fields as spatially homogeneous. 
The spatially inhomogeneous transverse profile of LG beam as shown in Fig.\ref{Figure2i}a give rise to inhomogeneous character of absorption ($\textrm{Im}\langle\rho_{41}\rangle$). 
However the bright and dark regions of the coherence get interchanged as compared to the spatial inhomogeneity of LG beam.
This formation can be explained from Eq. (\ref{eq:perturbation}), where $\textrm{Im}\langle\rho_{41}\rangle \propto -\imath\Omega_1$ at resonance condition.
Fig.\ref{Figure2i}b shows the phase singularities of the LG beam which are located at $\phi=0$ and $\phi=2\pi$, and are mapped to the phase of the induced coherence $\langle\rho_{41}\rangle$ structure as depicted in Fig.\ref{Figure2i}d.
We observed a extra rotation of $3\pi/2$ developed in the induced phase singularity that comes from the $e^{3\i\pi/2}$ factor that appear in the numerator of the Eq. (\ref{eq:perturbation}).
Hence the characteristic features of absorption and phase of the induced coherence are the essence of singular beam generation.

%**************************************************************************************************************************% 
%**************************************************************************************************************************%
Next, we derive the conditions for successfully mapping  spatial structure of phase as well as intensity profile of the light beams to the generated FWM signal.
The faithful generation of signal  is possible by satisfying the phase matching conditions and fulfilment of OAM conservation laws as well as the Gouy phase matching criterion.
A collinear beam geometry, where the proper choice of propagation direction of the interacting fields
can secure the phase-matching condition, {\it i.e.,} $\vec{k}_{g}=\vec{k}_{1}+\vec{k}_{2}-\vec{k}_{3}$, is used in our study.
The coherent FWM process \cite{Pruvost2018} inherently executes OAM conservation law {\it i.e.,} $l_{g}+l_{3}=l_{1}+l_{2}$.
The Gouy phase matching criterion is another important requirement which immensely controls the nonlinear frequency conversion process
\cite{Pruvost2018}.
The Gouy phase appears in the OAM carrying light beams as $\psi(m,l,z)=(2m+|l|+1)\tan^{-1}(z/z_0)$.
The efficient conversion confirms that the Gouy phase of the applied vortex beams should be equal to the generated FWM signal at any propagation distance $z$.
We first choose all the fields in resonance with their respective atomic transitions.
With all these stringent conditions, the individual OAM of probe beam $(l_{1})$ as well as control beam $(l_{2})$ or
a combination of $l_{1}$, $l_{2}$ can be cloned from optical beams into the generated signal.
%###############################################################################%
\begin{figure}
\includegraphics[scale=0.72]{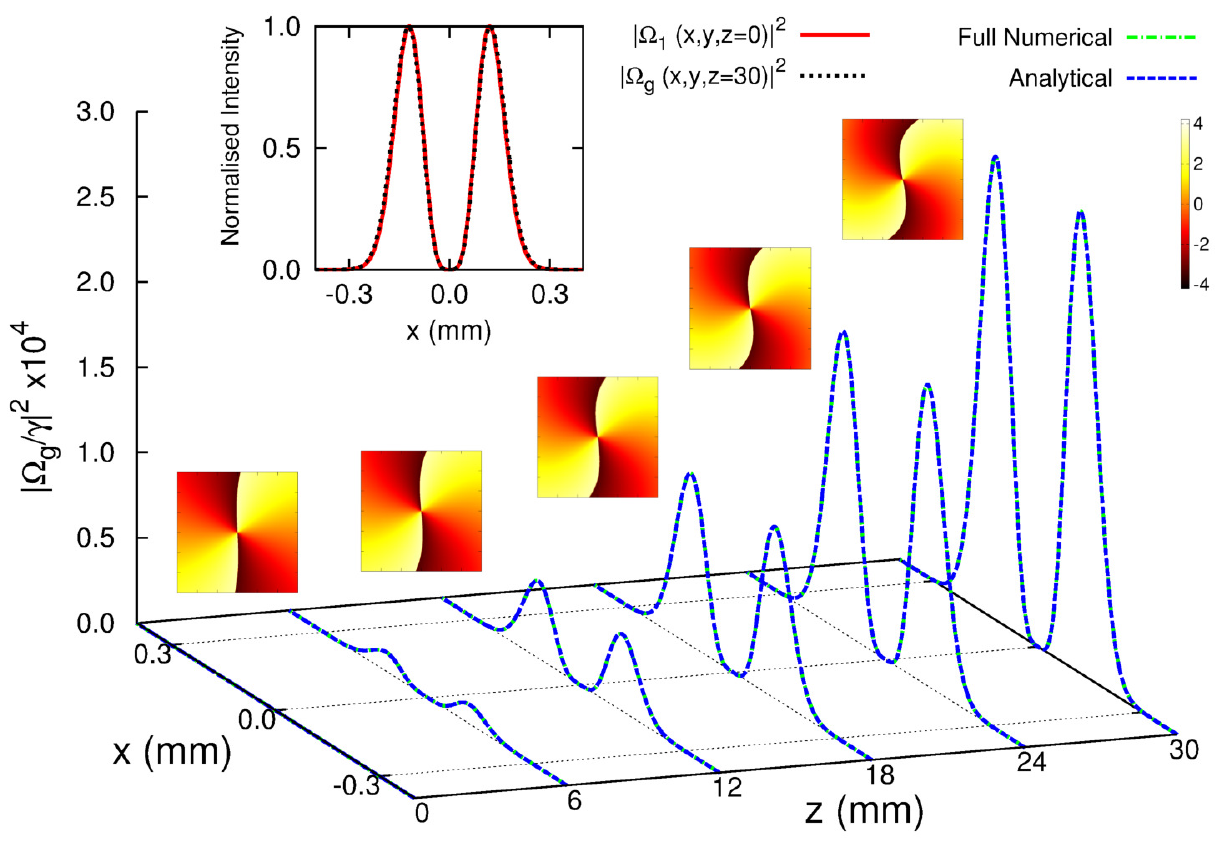}
\caption{The evolution of the FWM signal ($\Omega_g$) and phase structure.
Green dashed dot line (blue dashed line) shows the full numerical (analytical) dynamics of $\Omega_g$.
Inset figure compares the normalised intensity profile of the output $\Omega_g$ and input $\Omega_{1}$.
The parameters are $m_{1}=0$, $l_{1}=2$, $w_{1}=120\mu m$, $\Omega^{0}_{1}=0.01\gamma$, $\Omega^{0}_{2}=0.1\gamma$,
$\Omega^{0}_{3}=1.0\gamma$, $\Delta_{21}=0$, $\Delta_{32}=0$, $\Delta_{34}=0$, T=300K.}
\label{Figure3}
\end{figure}  
%###############################################################################%

We numerically solve Eq. (\ref{propagation}) using the split-step Fourier method (SSFM) for the progression
of the FWM signal to confirm successfully mapping the induced coherence $\langle\rho_{41}\rangle$ to signal. 
The Doppler averaged nonlinear atomic coherence $\langle\rho_{41}\rangle$
in Eq. (\ref{propagation}) governs that various OAM conversion processes.
Both the intensity and phase profile of the FWM signal can be pondered for confirmation of its actual vortex mode.
The spatial dynamics of the probe and control beams can be neglected as it does not affect the progression of the FWM signal.  
%**************************************************************************************************************************%
\begin{figure}[b]
\includegraphics[scale=0.9]{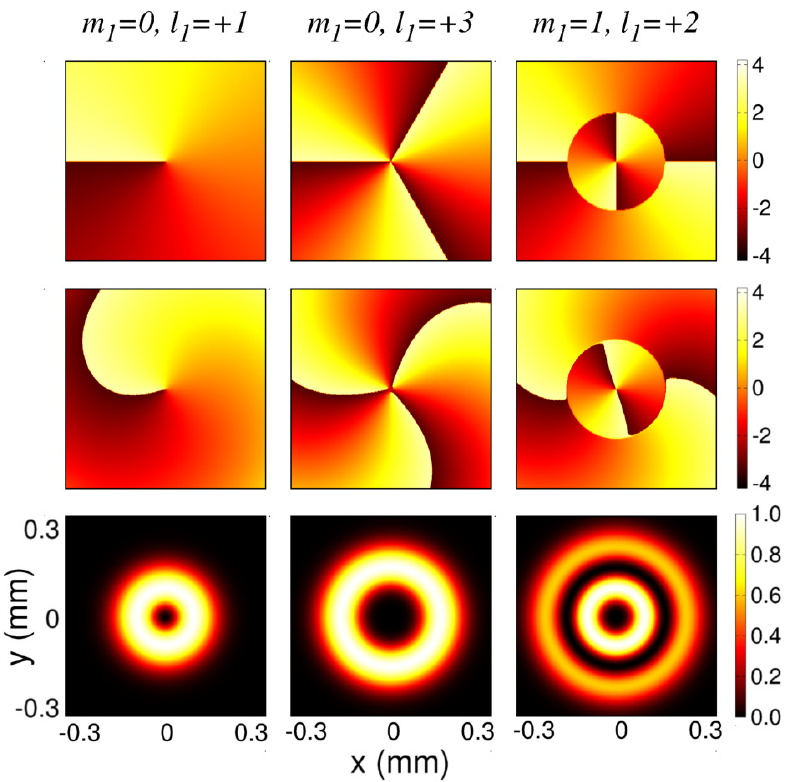}
\caption{Transfer of different OAM from probe beam ($\Omega_1$) to FWM signal ($\Omega_g$).
First row shows phase profile of $\Omega_1$ due to different OAM at $z$=0.
Second and third row depict the phase and intensity profile of $\Omega_g$ at $z$=30mm.
Other parameters are same as shown in Fig. \ref{Figure3}.}
\label{Figure4}
\end{figure}
%**************************************************************************************************************************%
%###############################################################################%
%**************************************************************************************************************************%
\begin{figure}[t]
\includegraphics[scale=0.3]{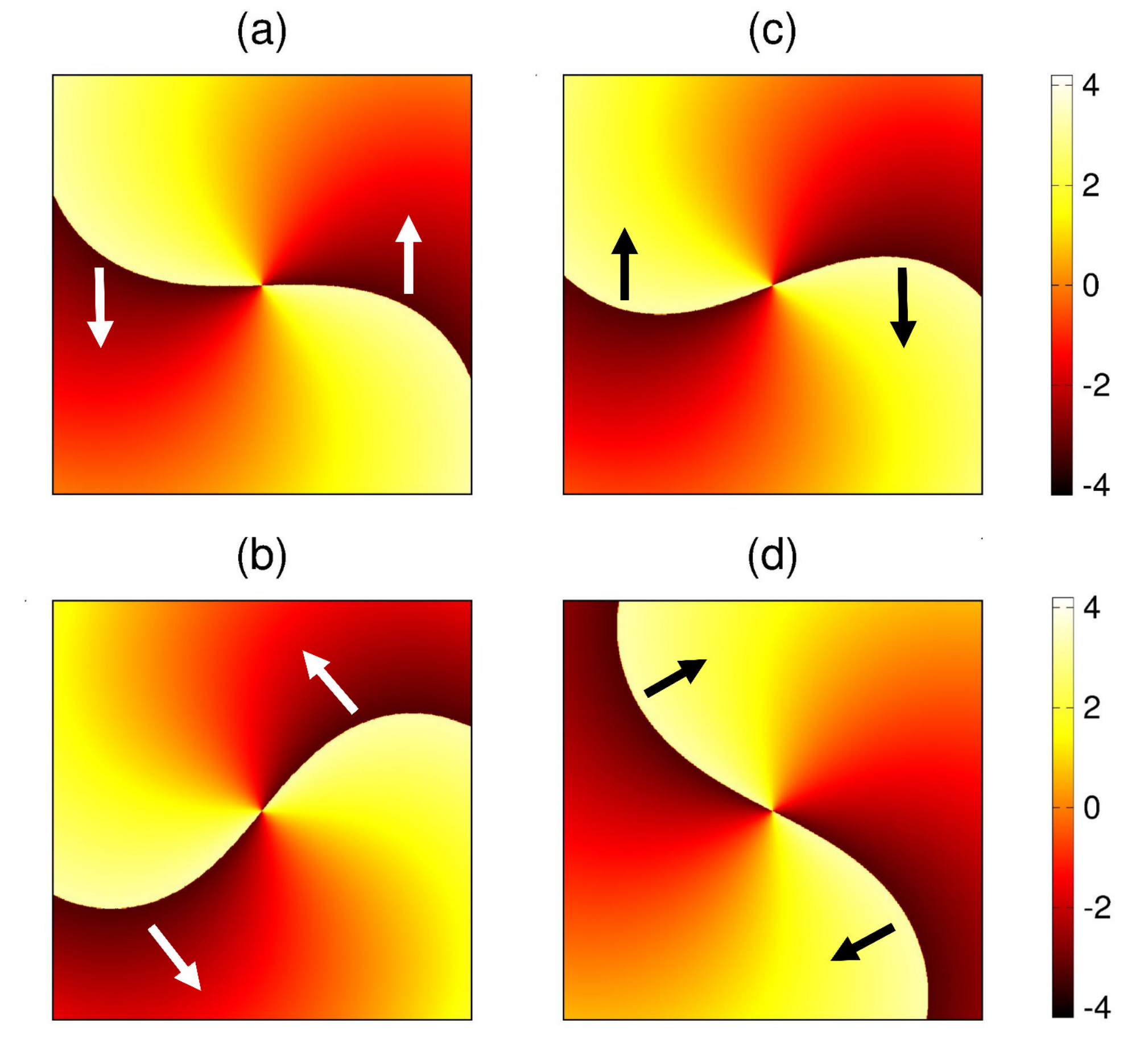}
\caption{Phase rotation of the FWM signal at z=30 mm for different values of probe detuning, $\Delta_{21}$ under two-photon
resonance condition, $\Delta^{'}_{21}$+$\Delta^{'}_{32}$=0. White arrows in
(a) $\Delta_{21}=-1.0\gamma$, (b) $\Delta_{21}=-10.0\gamma$ show anti-clockwise rotation and black arrows in (c) $\Delta_{21}=1.0\gamma$,
(d) $\Delta_{21}=10.0\gamma$ show clockwise rotation with respect to the input phase profile of $\Omega_1$.
Other parameters are same as shown in Fig. \ref{Figure3}.}
\label{Figure5}
\end{figure}
%**************************************************************************************************************************%         
%###############################################################################%
\subsection{Transfer OAM of $\Omega_1$}
%###############################################################################%
%**************************************************************************************************************************% 
\begin{figure}[b]
\begin{center}
\includegraphics[scale=0.7]{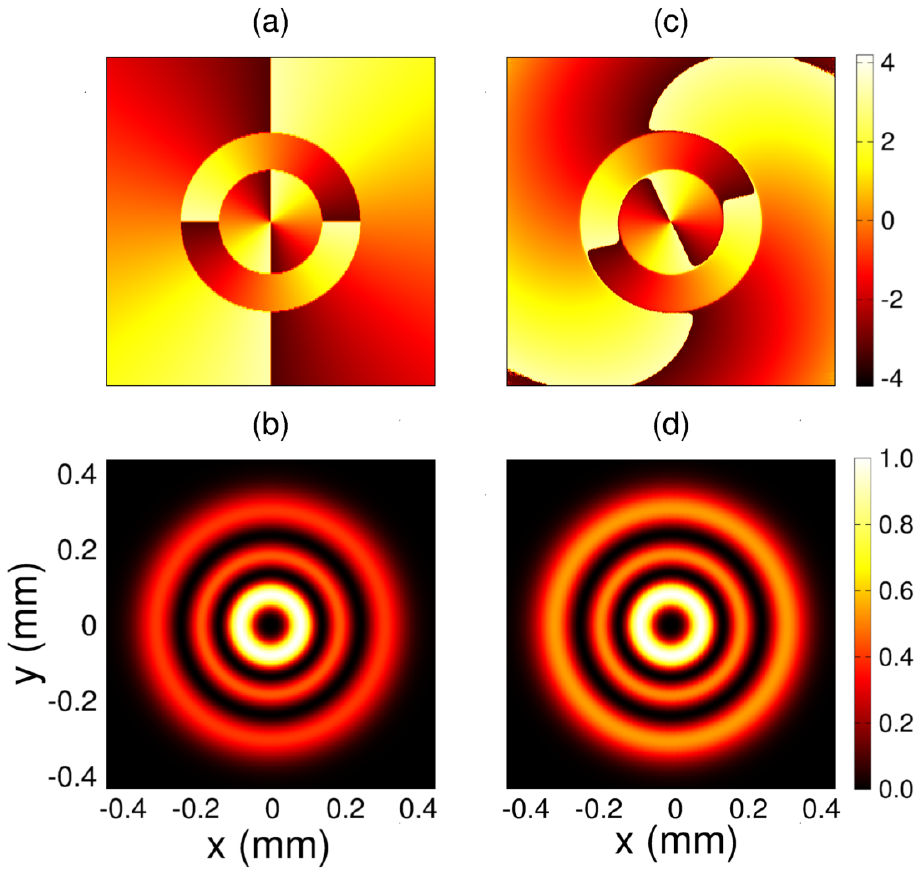}
\includegraphics[scale=1.0]{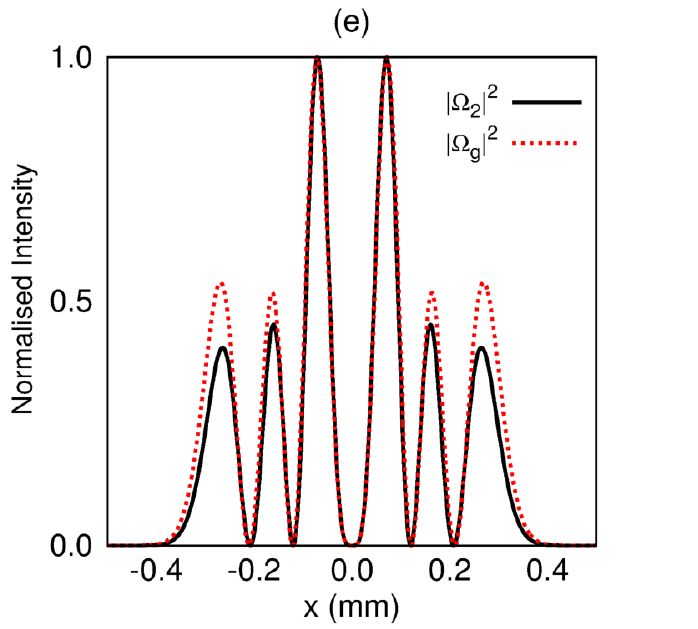}
\caption{Transfer of control beam's OAM ($l_{2}$) is demonstrated. Input phase and intensity profile of the control beam [(a) and (b)].
Output phase and intensity profile of the FWM signal [(c) and (d)].
(e) Comparison of normalised intensity profile of input control beam and output FWM signal.
The parameters are $m_{2}=2$, $l_{2}=2$, $w_{2}=120\mu m$. Other parameters are same as shown in Fig. \ref{Figure3}.}
\label{Figure6}
\end{center}
\end{figure}
%**************************************************************************************************************************%
We now exhibit our results by considering different transverse shape of the probe beam ($\Omega_1$) by selecting various values of $l_{1}$ and $m_1$  of the Laguerre-Gaussian mode. 
Other two fields $\Omega_2$, $\Omega_3$ are chosen to be continuous wave (cw) as its gives a plane wave front.
In the first step, we focus on generation of FWM signal that gets cloned from the vortex probe beam with mode $m_{1}=0$ and $l_{1}=+2$. 
In Fig.\ref{Figure3}, we have plotted the transverse variation of the generated intensity $|\Omega_g/\gamma|^2$ at different propagation lengths of the medium together with azimuthal phase.
In the inset of Fig.\ref{Figure3}, we show the agreement between the two different numerical methods that ensures the reliability of our results.
These results are obtained from the beam propagation equation (\ref{propagation}) by using coherence $\langle\rho_{41}\rangle$ derived from matrix inversion of the full set of the density-matrix equations (\ref{eq:dynamical}),  at steady state limit and the analytical perturbation expression (\ref{eq:perturbation}).
This inset also confirm that the generated signal at $z$=30 mm acquire identical intensity profile of the input probe beam as indicated by Eq. (\ref{eq:perturbation}), $\rho_{41} \propto -\imath\Omega_{1}$ \cite{Akulshin2015}.
We also notice that the spreading of generated beam due to diffraction is insignificant since its Rayleigh length, $z_{g}=\pi w^{2}_{g}/\lambda$ is much larger than the propagation distance $z$.
As seen from Fig.\ref{Figure3}, the generated phase experiences a rotation in the transverse plane as it propagates along $z$-direction which is an inherent feature of a vortex beam.
The phase structure of the FWM signal carries two distinct singularities at 0 and 2$\pi$ location.
A phase change of 4$\pi$ establishes the fact that the FWM assisted optical vortex has same features as the input probe vortex.
Next, a representative set of input probe vortex mode {\it e.g.}, $m_{1}=0, l_{1}=1$; $m_{1}=0, l_{1}=3$ and $m_{1}=1, l_{1}=2$;
is taken into consideration for our analysis in order to justify the robustness of the proposed system.
In Fig. \ref{Figure4}, the first row represents the phase structure of the input probe beam carrying different OAM.   
Whereas the second and third row of Fig. \ref{Figure4} depict the corresponding phase
and intensity profile of the generated field at a propagation distance of $z$=30 mm.
It is evident from Fig. \ref{Figure4} that the OAM of the probe beam precisely transfer into the generated FWM signal.

Next we discuss how the rotation of the generated phase front can be manipulated by changing the probe beam detuning $\Delta_{21}$ under two-photon resonance condition. 
We maintain the infrared field on resonance {\it i.e.,} $\Delta^{'}_{32}=0$, whereas control and probe fields fulfil two-photon resonance condition {\it i.e.,} $\Delta^{'}_{21}+\Delta^{'}_{32}=0$. 
A additional phase factor due to single photon detuning $\Delta_{21}$ appear in the phase structure of generated LG beam in the form of $e^{ \mp i\phi_{d}}$.
This detuning induced phase structure, $\phi_{d}$ can be defined as
%%%%%%%%%%%%%%%%%%%%%%%%%%%%%%%%%%%%%%%%%%%%%%%%%%%
\begin{equation}\label{phase}
\phi_{d}=\tan^{-1}\left(-\frac{\Delta^{'}_{21}}{\frac{\gamma_{2}+\gamma_{1}}{2}+\frac{|\Omega_2|^2}{\frac{\gamma_{3}+\gamma_{1}}{2}+\frac{|\Omega_3|^2}{\frac{\gamma_{4}+\gamma_{1}}{2}}}}\right).
\end{equation}
%%%%%%%%%%%%%%%%%%%%%%%%%%%%%%%%%%%%%%%%%%%%%%%%%%%
In Fig. \ref{Figure5}, the phase front of the output FWM signals $(m_{1}=0$, $l_{1}=+2)$
are plotted for different values of probe detuning $\Delta_{21}$, ranging from red to blue detune.
It is clearly observed from Fig. \ref{Figure5} that the polarity and magnitude of the detuning plays a key role in controlling the amount and direction of phase rotation of the generated field.

%###############################################################################%
\subsection{Transfer OAM of $\Omega_2$}
%###############################################################################%
We now investigate transfer of OAM from control beam ($\Omega_2$) to FWM signal.
For this purpose, the transverse profile of the control field is taken to be a higher-order Laguerre-Gaussian mode with $m_{2}$=2, $l_{2}$=2, whereas
probe and infrared field $\Omega_1$, $\Omega_3$ are chosen as cw for plane wave-front.
Fig. \ref{Figure6}(a) and \ref{Figure6}(b) delineate the phase structure and intensity
profile of the input control beam. 
The output phase and intensity profile of the FWM signal at $z$=30 mm are shown in
Fig. \ref{Figure6}(c) and \ref{Figure6}(d).
In order to check the fidelity of the conversion, we compare the normalised intensity
profile of $\Omega_{2}$ at $z$=0 and $\Omega_{g}$ at $z$=30 mm in Fig. \ref{Figure6}(e). 
This study confirms that the FWM signal and control beam
have similar intensity profile and phase structure. Note that
the transfer of OAM induced information from the control beam into the FWM signal
can be achieved only when $\Omega_2 < \Omega_3$ as suggested by Eq. (\ref{eq:perturbation}).
Under this condition, the term, $|\Omega_2|^2/[\Gamma_{31}+|\Omega_3|^2/\Gamma_{41}]$
in the denominator of Eq. (\ref{eq:perturbation}) can be neglected and phase and amplitude informations imprinted on the control beam
can be profoundly mapped into the generated signal with high fidelity. 

%###############################################################################%
\subsection{Transfer OAM of $\Omega_1$ and $\Omega_2$}
%###############################################################################%
%**************************************************************************************************************************%
\begin{figure}[t]
\includegraphics[scale=0.8]{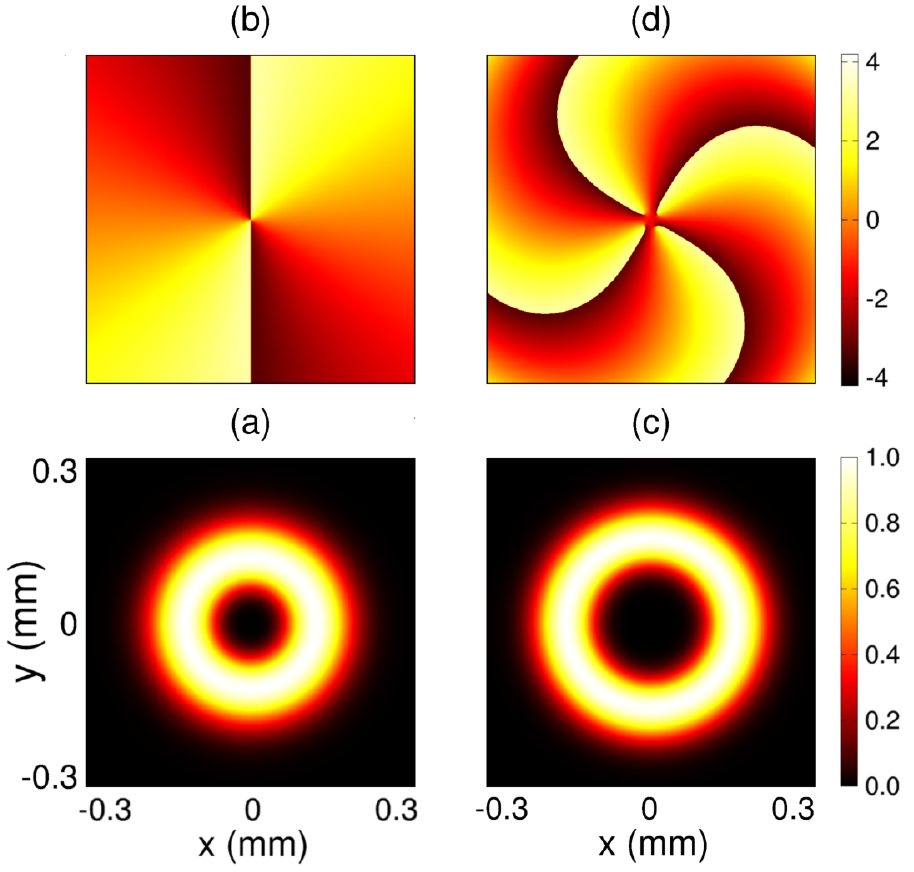}
\caption{Simultaneous transfer of probe OAM ($l_{1}=2$) and control OAM ($l_{2}=2$) into the FWM signal such that $l_{g}=l_{1}+l_{2}$.
Normalised intensity and phase profile of the input probe and control beam [(a) and (b)].
Output intensity and phase profile of the FWM signal [(c) and (d)].
The parameters are $m_{1}=0$, $m_{2}=0$, $l_{1}=2$, $l_{2}=2$, $w_{1}=120\mu m$, $w_{2}=120\mu m$.
Other parameters are same as shown in Fig. \ref{Figure3}.}
\label{Figure7}
\end{figure}
%**************************************************************************************************************************%  
In this section, we explore simultaneous transfer of probe and control beam's OAM into the generated FWM signal.
We consider both the optical beams ($\Omega_1$ and $\Omega_2$) possess the Laguerre-Gaussian mode, $m_{j}=0, l_{j}=2$; $j\in \{1,2\}$
and $\Omega_3$ is chosen as cw with plane wave-front.
From vortex beam generation criterion, we have found that OAM induced phase structure of the output FWM signal is the sum
of individual probe and control beam's OAM {\it i.e.,} $l_{g}=l_{1}+l_{2}$.
Fig. \ref{Figure7}(a) and \ref{Figure7}(b) represent
the intensity and phase profile of the input probe and control beam, respectively.
The intensity and phase profile of the generated signal at $z=30$mm are depicted in Fig. \ref{Figure7}(c) and \ref{Figure7}(d).
These results support that FWM signal gets its shape from its generator beams and it obeys $\Omega_{g} \propto \Omega_{1}\Omega_{2}$. 
Subsequently higher OAM can be generated by using the sum rule of probe and control OAM, $l_{1}+l_{2}$, which 
is efficiently transferred into the FWM signal as it is clearly depicted in Fig. \ref{Figure7}(d).
It should be borne in mind that nonlinear atomic coherence ($\rho_{41}$) in Eq. (\ref{eq:perturbation}) can be proportional to $\Omega_{1}\Omega_{2}$
under the condition of $\Omega_2 < \Omega_3$. This plays an important role in the transfer of the OAM of probe and control beam simultaneously on the generated beam.        

%##########################################################################%
\section{CONCLUSION}
\label{CONCLUSION}
%##########################################################################%
In conclusion, we have demonstrated an efficient FWM based OAM translation process in a 
diamond-type inhomogeneously broadened $^{85}$Rb atomic system.
Two optical fields and one infrared field nonlinearly interact with
the atoms and initiate phase-matched non-degenerate FWM signal. In this nonlinear process, the vortex
modes imprinted on the probe or control beam are transferred into the generated FWM signal with high fidelity.
Also, additional rotation of the phase structure in both clockwise and anti-clockwise direction induced by the single photon detuning
is observed which has important application in optical tweezers.
The generation and manipulation of such optical vortices through nonlinear FWM processes
in atomic system can have promising applications in optical communication and quantum information processing systems.         

\section*{Acknowledgments}
N.S.M. would like to thank MHRD, Government of India for financial support. T.N.D. thanks Prof. G. S. Agarwal for discussions and much useful input and gratefully acknowledges funding by the Science and Engineering Board (Grant No. CRG/2018/000054).

\bibliography{reference}
\end{document}